\documentclass{agujournal2018}
\usepackage{apacite}
\usepackage{times}
\usepackage{float}
\usepackage{amsmath}
\usepackage{epsfig}
\usepackage{url}
\usepackage[breaklinks,colorlinks,citecolor=blue]{hyperref} 

\definecolor{mypurple}{rgb}{0.33,0.11,0.46}

\journalname{JGR: Planets}

\begin{document}


\title{Habitable Snowballs: Temperate Land Conditions, Liquid Water, and Implications for CO$_2$ Weathering}
\authors{Adiv Paradise\affil{1,2}, Kristen Menou\affil{1,2}, Diana Valencia\affil{1,2}, Christopher Lee\affil{3}}

\affiliation{1}{Department of Astronomy and Astrophysics, University of Toronto, St. George, Toronto, ON M5S 3H4, Canada}
\affiliation{2}{Centre for Planetary Sciences, Department of Physical and Environmental Sciences, University of Toronto, Scarborough, ON M1C 1A4, Canada} 
\affiliation{3}{Department of Physics, University of Toronto, St. George, Toronto, Ontario, M5S 1A7, Canada}

\correspondingauthor{Adiv Paradise}{paradise@astro.utoronto.ca}

\begin{keypoints}
\item Direct sunlight can warm bare land areas above freezing even if the rest of the planet is frozen
\item Warm temperatures and elevated CO2 lead to continental weathering during snowball events
\item Snowball planets with low outgassing rates may both be stable and have habitable land surfaces
\end{keypoints}




\begin{abstract}

Habitable planets are commonly imagined to be temperate planets like Earth, with areas of open ocean and warm land. In contrast, planets in snowball states, where oceans are entirely ice-covered, are believed to be inhospitable. However, we show using a general circulation model that terrestrial planets in the inner habitable zone are able to support large unfrozen areas of land while in a snowball state. Due to their lower albedo, these unfrozen regions reach summer temperatures in excess of 10 $^\circ$Celsius. Such conditions permit CO$_2$ weathering, suggesting that continental weathering can provide a mechanism for trapping planets in stable snowball states. The presence of land areas with warm temperatures and liquid surface water motivates a more-nuanced understanding of habitability during these snowball events.

\end{abstract}

\section*{Plain Language Summary}

Studies examining the ability of Earth-like planets to host life have often used conditions leading to `snowball' events, where sea ice extends all the way to the equator, as a limit to the range of habitable climates. This has been based on the assumption that snowball planets are always below-freezing everywhere on their surfaces. As the chemical process by which CO$_2$ is removed from the atmosphere and bound in surface rocks relies on warm temperatures and liquid water, and therefore would not happen in globally-cold conditions, this has also led to the conclusion that snowball events should be temporary, coming to an end when volcanoes release enough accumulated CO$_2$ to warm the planet enough to melt the ice. We ran several thousand three-dimensional computer simulations of Earth-like climates and found that if there are inland areas of dark, bare ground, under enough sunlight those regions can be warm enough for liquid water and life without causing the sea ice to retreat. This suggests that snowball planets should not be excluded as inhospitable to life, and that on some planets the burial of CO$_2$ in surface rocks in these areas could balance volcanic emissions, resulting in permanent snowball conditions.



%
%

\section{Introduction}
\subsection{Snowball Climates}

Earth-sized exoplanets are often characterized by their orbits relative to their star's `habitable zone', the region in which surface temperatures can support liquid water. This is usually determined using climate models, with the inner edge as the distance where planets transition to inescapably hot runaway greenhouses, such as Venus, and the outer edge as the distance where models freeze over even with increased greenhouse gases \citep{Hart1979,Kasting1993}. The inner and outer edges are often generalized to other parameters beyond distance by defining a `hot edge', a `soft cold edge' where sea ice reaches the equator, and a `hard cold edge' where carbon dioxide condenses \citep[e.g.][]{HaqqMisra2016,Abbot2016,Paradise2017}. The use of these parametric boundaries to describe the habitable zone often carries the implicit assumption that climates beyond the soft cold edge are globally cold.

The ice-covered planets beyond the soft cold edge are `snowball' planets, characterized by complete or very extensive sea ice coverage, a lower tropopause, strong Hadley circulation, and equatorial downwelling \citep{pierrehumbert05,Abbot2014,Hoffman2017}. Due to the high albedo of sea ice, these planets can remain frozen even at high levels of incident sunlight and greenhouse gas concentrations, resulting in a bistability \citep{Budyko1969,Sellers1969}. On fast-rotating planets such as Earth, the coupling between sea ice and surface temperature results in a positive feedback, leading to sharp transitions between snowball and temperate states \citep{Budyko1969,icealbedo}. Geological evidence suggests Earth has gone through snowball episodes possibly twice in its history \citep{Hoffman2002,dropstones,Tajika2007}. A return to temperate conditions requires global melting, triggered by substantially elevated greenhouse gases or large increases in insolation \citep{Budyko1969,hoffman98,pierrehumbert05}.

Snowball episodes are often considered to be problematic for life \citep[e.g.][]{pierrehumbert05,lucarini2013,HaqqMisra2016}, despite a lack of evidence of decreased biodiversity following Earth's snowball episodes, and even some evidence of an increase in biodiversity, though Earth life at this point was predominantly marine \citep{Corsetti2006}. However, it is not necessarily true that very low average global temperatures imply sub-freezing conditions everywhere. Low-dimensional models which deal primarily with global averages may not capture small regions with warmer temperatures. \citet{Spiegel2008} proposed that planets could be partially or temporally habitable (as in, only at certain times of the year), and \citet{Linsenmeier2015} showed that this is the case for planets with high obliquity or high eccentricity. \citet{Benn2015} and \citet{Hoffman2017} observed in coupled climate-cryosphere models that Marinoan Snowball experiments with high CO$_2$ levels could result in some land areas becoming dry and occasionally reaching warm temperatures. These models however focused on specific conditions in Earth's past. In contrast, we use an intermediate-complexity GCM to show that locally-temperate conditions could occur more generally in Earth-like planets near the snowball exit threshold. 

\subsection{Snowball Stability}

On Earth, the planet's surface temperature is regulated on geological timescales by the carbon-silicate cycle. Carbon dioxide (CO$_2$) in the air is dissolved into rainwater and delivered to the surface, where it undergoes a temperature-sensitive weathering reaction with rock to form carbonates. These rocks are eventually recycled into the mantle, where the CO$_2$ can be regassed into the atmosphere \citep{Sleep2001,Berner2004,pierrehumbertbook}. On Earth-like planets, the weathering rate is strongly coupled to the surface temperature through the reaction efficiency and the precipitation rate, so that weathering and outgassing form a negative feedback between the surface temperature and CO$_2$ level \citep{walker81,wk97,kump00,pierrehumbertbook}. Therefore, there is a global temperature at which outgassing and weathering are in equilibrium, and the climate will trend toward this temperature on geological timescales \citep{pierrehumbertbook}. 

On planets with low outgassing rates, this equilibrium temperature may be colder than the transition to a snowball state \citep{pierrehumbert05}. Previous studies have proposed that a cessation of weathering due to globally cold conditions could mean that such planets slowly outgas enough CO$_2$ to eventually exit snowball, resulting in limit cycles \citep{Menou2015,HaqqMisra2016,Paradise2017}. In other words, no stable point exists in low-outgassing regimes. However, if continental weathering were able to proceed during snowball episodes due to small seasonally-warm regions, this would provide an additional stable point for planets with low outgassing rates, allowing some planets to become trapped in snowball states. We therefore extend our methods in \citet{Paradise2017} to constrain the continental weathering rates of such climates, and therefore determine the limits of snowball stability.

\section{General Circulation and Weathering Model}
\subsection{PlaSim}

We use PlaSim, a 3D general circulation model (GCM) of intermediate complexity to simulate snowball climates. PlaSim uses spectral transforms to solve for vorticity, temperature, divergence, and pressure, and includes a 50-meter mixed-layer slab ocean model, thermodynamic sea ice, a simple soil model, and a 10-layer model atmosphere. Radiation is implemented using a three-band model, with two shortwave bands and one longwave band, and includes gray cloud scattering and absorption by water, CO$_2$, and ozone. The hydrological cycle is modeled as a series of interconnected parameterized physical processes, including shallow and deep cumulus convection, moisture condensation based on computed saturation specific humidities, large-scale precipitation, re-evaporation of falling rain and snow, and surface water storage, evaporation, and runoff \citep{Fraedrich2005}. In its T21 configuration, PlaSim has a resolution of $5.6^\circ\times5.5^\circ$ at the equator.

PlaSim has been used before to study snowball events on Earth-like planets \citep[e.g.][]{lucarini2010,Boschi2013,Linsenmeier2015}, and is able to reproduce snowball phenomena found using other models by \citet{pierrehumbert05}, \citet{Abbot2012b}, and \citet{Abbot2014}, such as increased variability in troposphere lapse rate, reduced tropopause height, weakened extratropical winter lapse rates, and warming of snowball states by cloud forcing. Models disagree on the extent of sea ice coverage during snowball events \citep[e.g.][]{Kirschvink1992,Lewis2007,Yang2012,Rodehacke2013}, but in PlaSim, snowballs are characterized by complete sea ice cover, and exit from a snowball event begins when sea ice starts to melt, triggering an albedo feedback runaway. Sea ice thickness is limited to 9 meters, but we do not restrict the depth of land ice and snow. Soil hydrology is described by a bucket model, where each grid cell has a prescribed water capacity. Excess surface water is treated as runoff and advected away according to the local mean topographical slope, forming PlaSim's river system \citep{Fraedrich2005}. Evaporation is driven by both runoff and ground water, and the presence of surface liquid water, snow, or ice changes the surface heat capacity.  We run models until the surface and top-of-atmosphere energy balance both change by less than 0.5 mW/m$^2$ per year over a 30-year baseline, which in most cases is 100-300 years. 

We assume the same land configuration as modern Earth, along with modern obliquity and eccentricity. As in \citet{Paradise2017}, we sample a range of insolations ranging from 1400 W m$^{-2}$ to 1050 W m$^{-2}$, as well as a range of CO$_2$ partial pressures (hereafter pCO$_2$), holding the partial pressure of the rest of the atmosphere constant, so that the surface pressure varies with pCO$_2$. We use both warm-start initial conditions (temperate conditions similar to modern Earth) and cold-start initial conditions (snowball conditions with a global annual average surface temperature near 220 K, with no above-freezing surface conditions anywhere at any time), increasing our sampling resolution near the snowball transition points. We measure the global weathering rate at each point. The models that equilibriate in snowball conditions just before the transition to globally warm models represent the warmest snowballs in our sample, and which we therefore assume to represent the snowball exit threshold. We focus on the warm end of the snowball regime because that represents the upper limit on warm temperatures and continental weathering possible in snowball climates, and CO$_2$ outgassing will tend to push colder snowball climates toward this warmer regime.

\subsection{Weathering Model}

We implement weathering at each gridpoint following the technique in \citet{Paradise2017}, itself similar to techniques used in \citet{Lehir2009} and \citet{Edson2012}. In \citet{Paradise2017}, however, we parameterized precipitation rates in terms of surface temperature, rather than treating the role of precipitation directly. Here, we make use of the GCM's grid-level data for precipitation to include it directly. Therefore, our weathering parameterization is given as 
\begin{linenomath*}
\begin{equation}
\frac{W}{W_\oplus} = \kappa\left(\frac{pCO_2}{{pCO_2}_\oplus}\right)^\beta\left(\frac{\phi}{\phi_\oplus}\right)^{0.65}e^{[k_\text{act}(T_\text{s}-288)]}\label{eq:weathering}
\end{equation}
\end{linenomath*}
where $W_\oplus$ is the weathering rate assumed for modern Earth, $k_\text{act}$ is related to the reaction's chemical activation energy and is set to 0.09 \citep{Berner1994,Berner2001}, and $T_\text{s}$ is the surface temperature. While in \citet{Berner1994}, $\phi$ is runoff, PlaSim concentrates runoff near continental margins, and so precipitation may be more useful for estimating local weathering. Total runoff is mostly proportional to total precipitation in Plasim, so following \citet{Abbot2012a}, we assume non-dimensional precipitation and runoff are interchangeable. We use 79 cm yr$^{-1}$ for the annual rainfall on land, $\phi_\oplus$ \citep{Chen2002,Schneider2014}. $\beta$ is set to 0.5 \citep{Berner1994}. Thus, the weathering increases if the amount of CO$_2$ increases, rainfall increases, or the available thermal energy increases. We note that the original parameterization is intended to be a function of average quantities, which is not necessarily equivalent to the average of the same function evaluated with local quantities. Therefore, we add a tuning factor $\kappa$, set to 11.9, to account for the deviation from modern Earth's weathering rate given modern Earth conditions. This prefactor also captures the effect of land fraction and the planet's radius, which we hold constant in this study. $\kappa$ is experimentally determined such that at 1367 W m$^{-2}$, surface pressure of 1 atm, and 360 ppmv CO$_2$ (PlaSim's default configuration for modern Earth), $W=W_\oplus$. 

The weathering rate is computed at each cell 4 times per day, which is frequent enough to capture daily temperature and rainfall variations, but avoids adding computational cost to every timestep (32 per day in this case). To compute the global weathering rate, we take an annual average for each cell and then compute an area-weighted average over the entire land surface. While outgassing rates may change slowly over time as the planet cools \citep{Zhang1993}, we simply assume a stable outgassing rate in equilibrium with modern Earth weathering, in line with previous work \citep[e.g.][]{Lehir2009,Edson2012,Menou2015}. Where relevant, we assume that Earth's modern steady-state weathering is 50 bar Gyr$^{-1}$ \citep{Gerlach2011,MartyTolsikhin98}. We use these units because we are directly interested in the change in atmospheric CO$_2$, rather than geological mass fluxes. 10 bar Gyr$^{-1}$ is approximately 1.2 Tmol C yr$^{-1}$, such that modern Earth weathering is 6 Tmol C yr$^{-1}$.

\section{Results}

\begin{figure}
\centering
\includegraphics[width=6in]{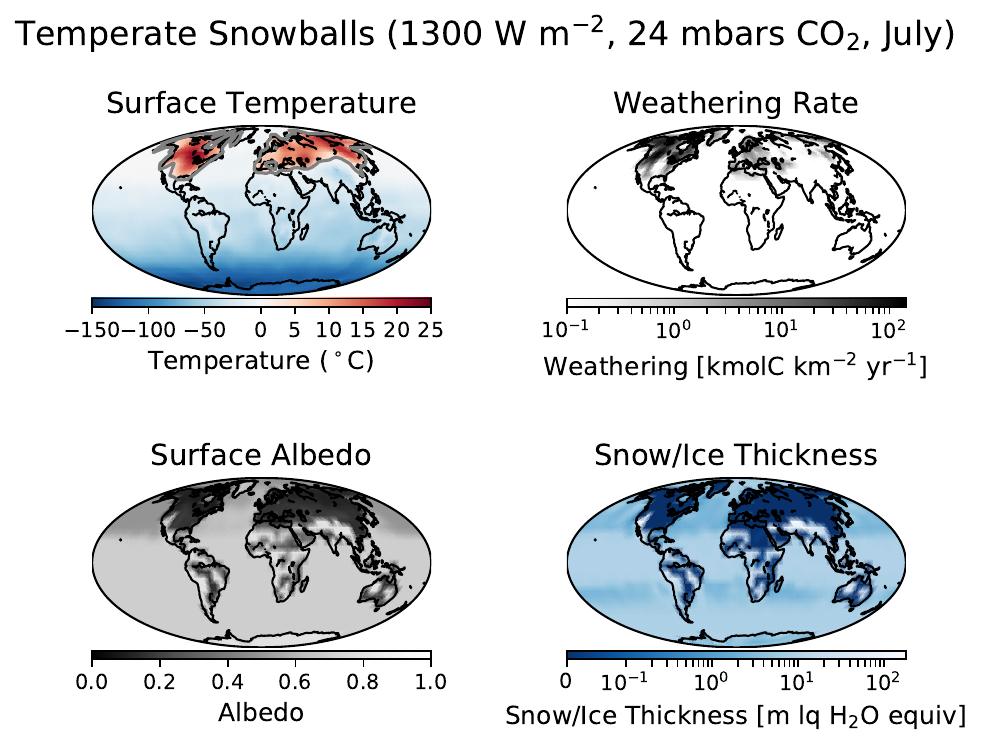}
\caption{Average July surface temperature, weathering rate, surface albedo, and snow/ice depth for a snowball planet with Earth-like continents and obliquity at 1300 W m$^{-2}$, or approximately 1.025 AU, with 24 mbars of CO$_2$. This is almost the CO$_2$ level needed to exit snowball at this insolation. Temperate regions, delineated by the gray 0 $^\circ$C isotherm, experience peak temperatures around 25 $^\circ$C and average summer temperatures above 10 $^\circ$C. This is the result of lower-albedo bare ground which absorbs more shortwave radiation than snow or ice. Rainfall over these regions during the summer enables significant weathering. This particular case has approximately 9\% of modern Earth weathering in total, and 35\% of the land is temperate in the summer.}\label{fig:localmaps}
\end{figure}

\begin{figure}
\centering
\makebox[\textwidth][c]{\includegraphics[width=1.3\textwidth]{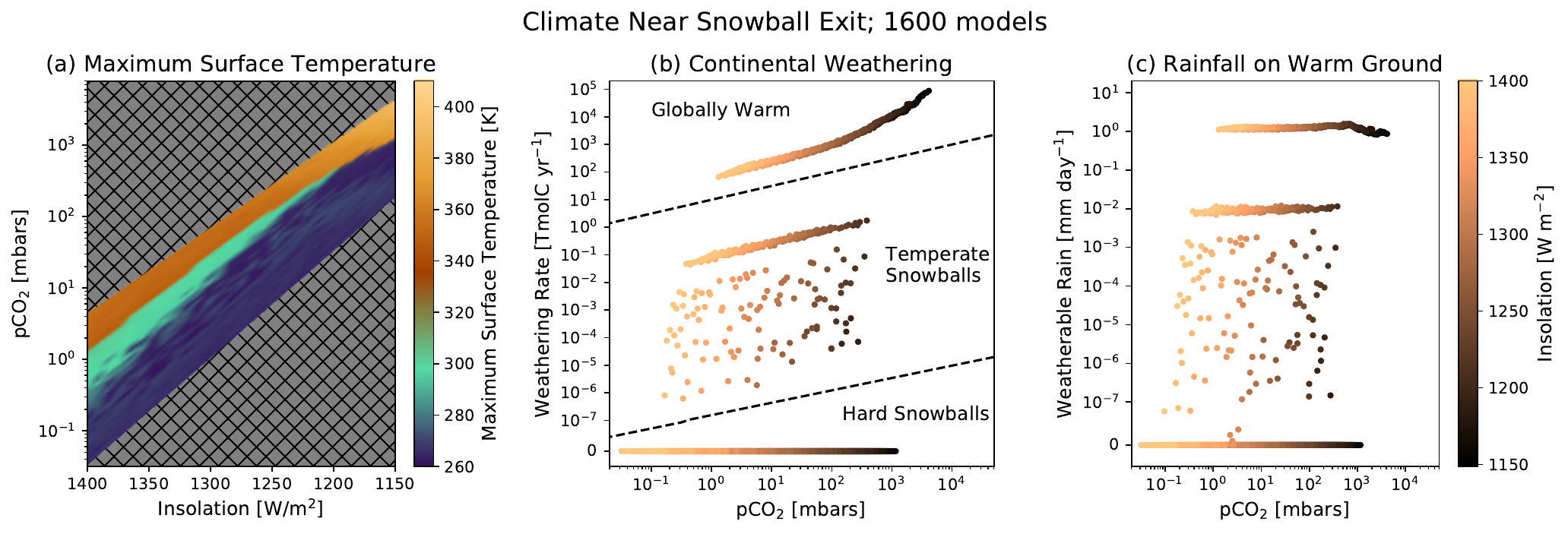}}
\caption{Annual maximum temperatures, net continental weathering, and precipitation for a grid of 1600 PlaSim models sampled across insolation and pCO$_2$, centered on the snowball exit threshold. Panel \textbf{(a)} shows the annual maximum surface temperature attained by PlaSim models across insolation and pCO$_2$. The green slice represents models where the planet is still in snowball, but maximum temperatures are in excess of 280 K. Above this regime are models that are globally warm, and below it are models in ``hard snowball", where temperatures never rise above freezing anywhere. The hatched gray background indicates parameter space excluded from this experiment. Panel \textbf{(b)} shows the continental CO$_2$ weathering calculated in each model from Panel \textbf{(a)}. Continental weathering is neatly segregrated into 3 regimes, corresponding to globally-warm conditions, temperate snowballs, and hard snowballs (zero weathering). Panel \textbf{(c)} shows the annual average precipitation onto warm ground---this is the only precipitation that is available for weathering.}\label{fig:parameterspacetemps}
\end{figure}

We find that snowball climates at moderate to high insolation and at the warmer end of the snowball regime are consistently characterized not just by complete sea ice coverage, but also by large ($>$100 km) regions of continental land that may undergo seasonal melting, attaining summer monthly average temperatures in excess of 10 $^\circ$Celsius. Warming is accompanied by increases in precipitation, driven by evaporation of meltwater and sublimation from ice surfaces. \autoref{fig:localmaps} illustrates this behavior in a specific Earth-like model with Modern Earth continents and axial tilt, 24 mbars of CO$_2$ (approximately 60 times present levels \citep{kunzig2013climate}), and 1300 W m$^{-2}$ insolation. This particular model has temperate conditions across 35\% of its land surface during northern hemispheric summer. These warmer temperatures are found both at the surface and in the lower few layers of the model's atmosphere. These temperate snowballs occur across an order of magnitude in pCO$_2$ at Earth-like insolations, but span much less at lower insolations, as shown in \hyperref[fig:parameterspacetemps]{Figure~\ref*{fig:parameterspacetemps}a}. Despite the relatively narrow parameter space, in the absence of other weathering mechanisms it may be more common to find snowball planets in this regime than at colder CO$_2$ levels, as net outgassing will cause colder snowballs to warm into the temperate snowball regime.

\begin{figure}
\centering
\includegraphics[width=6in]{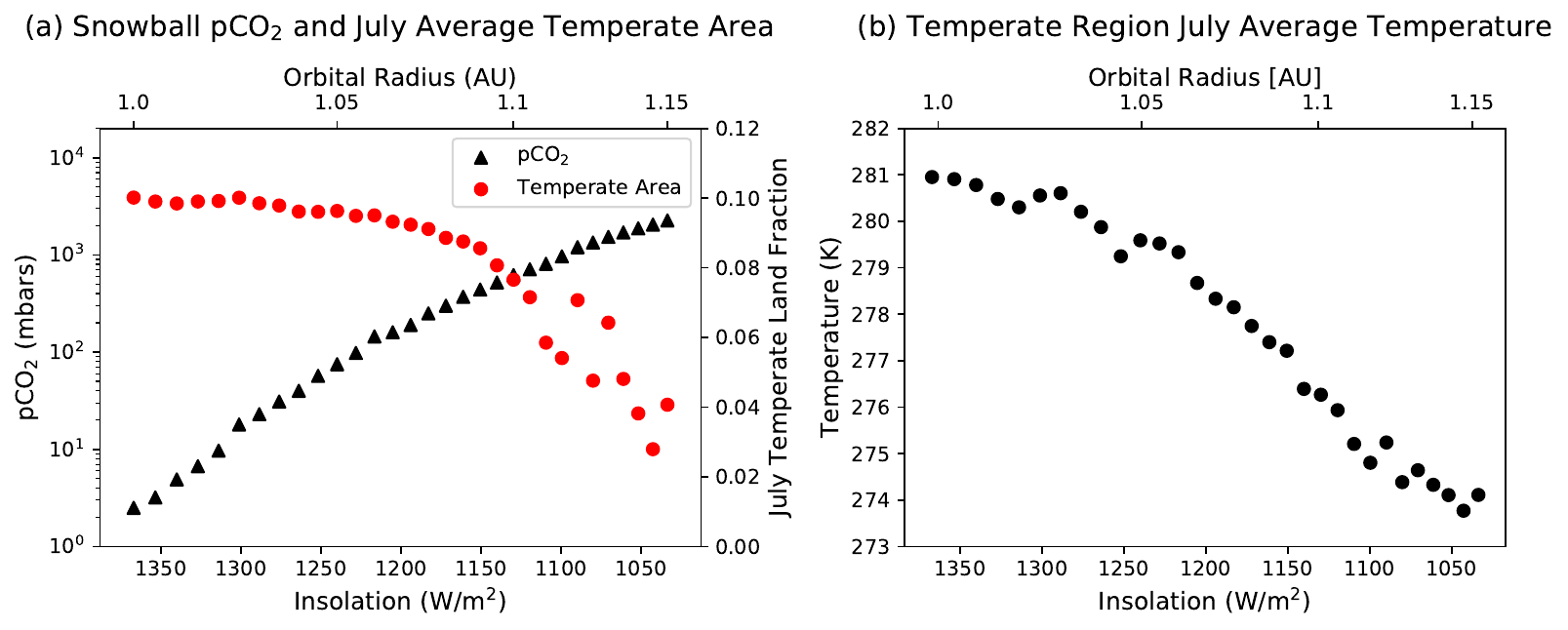}
\caption{\textbf{(a)}: The CO$_2$ partial pressure of snowball models just below the snowball exit threshold for a range of insolations (black), juxtaposed against the fraction of the planet's surface that is temperate during the summer (red). \textbf{(b)}: The average July surface temperature in regions which experience above-freezing conditions. Despite increased forcing from CO$_2$ absorption, temperate regions become smaller and cooler at lower insolations. This is because the shortwave response of bare soil is the controlling mechanism for the observed warming.}
\label{fig:trends}
\end{figure}

In addition to the permissible pCO$_2$ range for temperate conditions shrinking with decreasing insolation, the temperate regions themselves become both smaller and cooler at lower insolations, as shown in \autoref{fig:trends}. At lower insolations, where planets receive less shortwave flux, most of the climate forcing is due to CO$_2$ longwave forcing, while at higher insolations, lower pCO$_2$ leads to reduced longwave forcing and an energy budget dominated by shortwave forcing. This suggests that the dominant mechanism responsible for the warming is the strong shortwave response of bare ground compared to snow and ice, due to the lower shortwave albedo of bare ground. At lower insolations, despite similar net forcing, the weaker shortwave forcing results in less warming.

\begin{figure}
\centering
\includegraphics[width=4in]{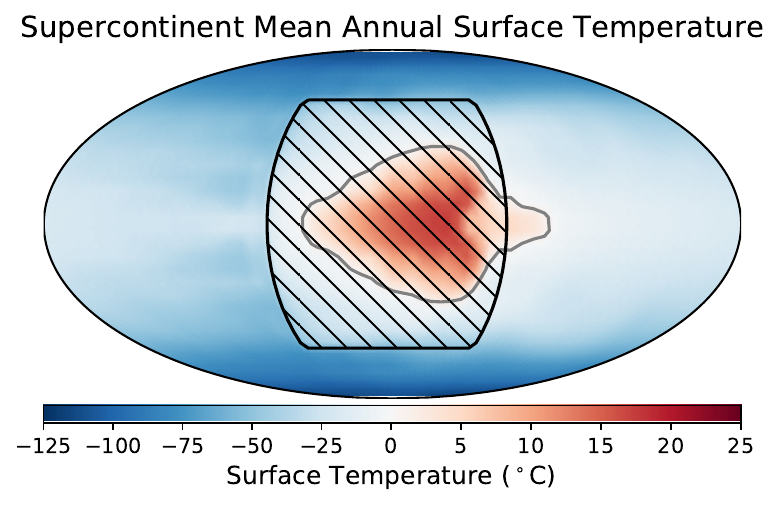}
\caption{Annual average surface temperature for a planet at 1300 W m$^{-2}$, with a hypothetical flat supercontinent (shaded rectangle) centered on the equator and pCO$_2$ below the snowball exit threshold. This planet has approximately 2.6 mbar of pCO$_2$ and a global annual average surface temperature of 242 K. However, approximately 41\% of the land surface boasts temperate annual average surface temperatures, permitting approximately 7.5\% of modern Earth weathering. Some snowball planets may therefore have regions with year-round temperate conditions. Snowball climates with modern Earth continents have annual average temperatures below 0 $^\circ$C everywhere.}\label{fig:pangea}
\end{figure}

We explore cases which might yield optimal habitability by considering the more favorable conditions of perpetual equinox (zero obliquity) and a large equatorial land mass. The equatorial regions of the land mass have solstice-like shortwave forcing year-round, potentially permitting persistent temperate regions. This represents an edge-case for snowball habitability, and also serves as a test of whether strong shortwave forcing and low surface albedo are the key factors for producing temperate snowballs. We therefore repeat our experiment with a rectangular equatorial supercontinent with the same total land area as modern Earth, as shown in \autoref{fig:pangea}. We confirm that this configuration results in persistent temperate land conditions.

%
%

\begin{figure}
\begin{center}
\includegraphics[width=4in]{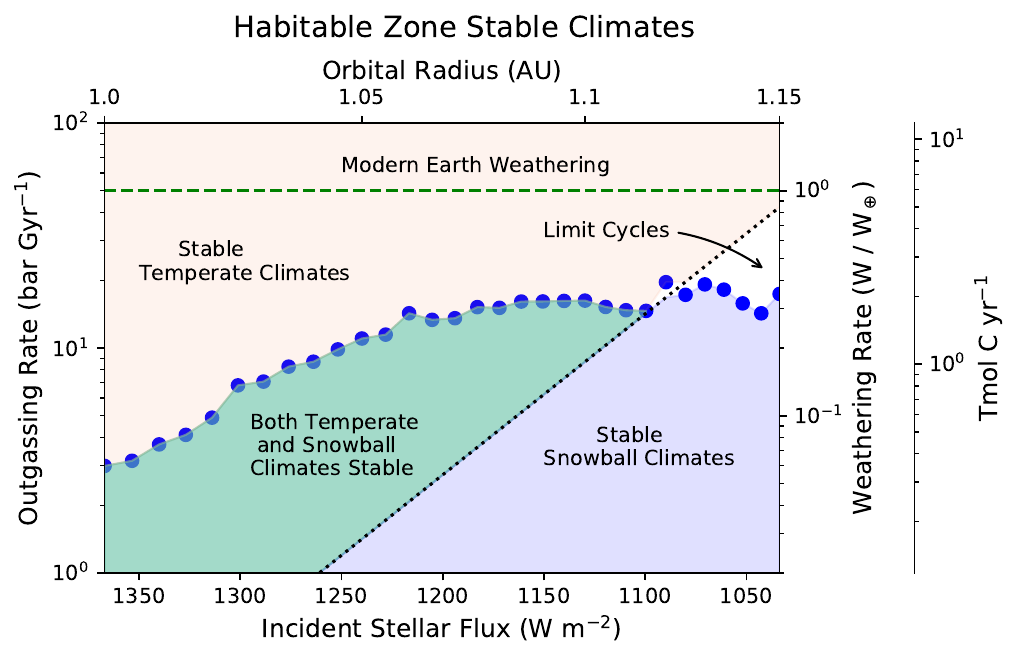}
\end{center}
\caption{The maximum weathering observed in snowball climate models just below the snowball exit threshold, depicted as blue circles. Continental weathering in transiently-temperate land regions during snowball represents an additional fixed point, as shown in panel \textbf{(b)} of \autoref{fig:parameterspacetemps}, permitting stable solutions for low outgassing rates, represented here as all areas below the blue circles. Globally-warm climates are stable for outgassing rates above the black dotted line, as found by \citet{Paradise2017}. Because the temperate snowball weathering limit and the globally-warm weathering limit intersect, there is a region of parameter space (green) where both climates are stable for a given outgassing rate.}\label{fig:weathering}
\end{figure}

We further find that temperate regions support substantial local continental weathering, as shown in the upper-right panel of \autoref{fig:localmaps}, resulting in moderate global weathering particularly at lower insolations, as shown in \autoref{fig:weathering}. At higher insolations, reduced pCO$_2$ leads to lower weathering rates, as shown in \hyperref[fig:parameterspacetemps]{Figure~\ref*{fig:parameterspacetemps}b}. This suggests that for planets with outgassing rates below a certain threshold, snowball states may be stable on geological timescales, implying that some fraction of observed terrestrial exoplanets may be in permanent snowball states. We speculate that such planets could be more mature, have cooler interiors, and less tectonic and volcanic activity \citep{Zhang1993}, or they could have formed with a smaller carbon inventory than Earth.

We note in \autoref{fig:weathering} that at low outgassing rates and high insolation, both states appear stable, in that the minimum weathering in the fully-temperate regime is less than the maximum weathering in the temperate snowball regime. This is a consequence of the very low pCO$_2$ at the cold end of the fully-temperate regime---even though average land temperatures are higher at the cold end of the fully-temperate regime than at the warm end of the snowball regime, and precipitation rates are accordingly higher, the minimum temperate pCO$_2$ is almost 1000 times less than the maximum snowball pCO$_2$. The precipitation dependence and CO$_2$ dependence are similarly important in \autoref{eq:weathering}, but precipitation at the warm end of the snowball regime is not 1000 times weaker than at the cold end of the fully-temperate regime. This results in the snowball entry point having lower weathering than the snowball exit point. Limit cycles once again become possible at low insolations because of the diminishing temperate regions.



\section{Discussion}

We find that in the inner habitable zone, the exposure of bare land during snowball episodes permits temperate conditions when incident shortwave radiation is strong, either seasonally for Earth-like continents and obliquity, or permanently for low-latitude land areas on zero-obliquity planets. The elevated pCO$_2$ levels and clement conditions permit continental silicate weathering which represents an additional stable point for low-outgassing planets. Because this result involves interactions with surface properties, the cryosphere, and the hydrological cycle, all of which are parameterized in PlaSim, we explore the sensitivity of our results to variations in physical properties such as ice sheet elevation, soil hydrology, erosion-motivated supply limits, and surface albedo. 


We find that weathering and temperate fraction vary by up to an order of magnitude in most sensitivity tests, and by many orders of magnitude when erosive supply limits and thick glaciers are considered. All of our sensitivity tests were performed independent of each other, so deviations in multiple parameters could increase the effects of the less-sensitive parameters. We also expect the parameterizations in PlaSim and in our weathering model to have biases, both known and unknown. These uncertainties however represent systematic biases, which means it is still possible to study the underlying physical processes. In this sense, our qualitative results are mostly insensitive to these parameters. More sophisticated models will however be necessary for quantitative predictions.


\subsection{Soil Albedo}

\begin{figure}
\begin{center}
\includegraphics[width=6in]{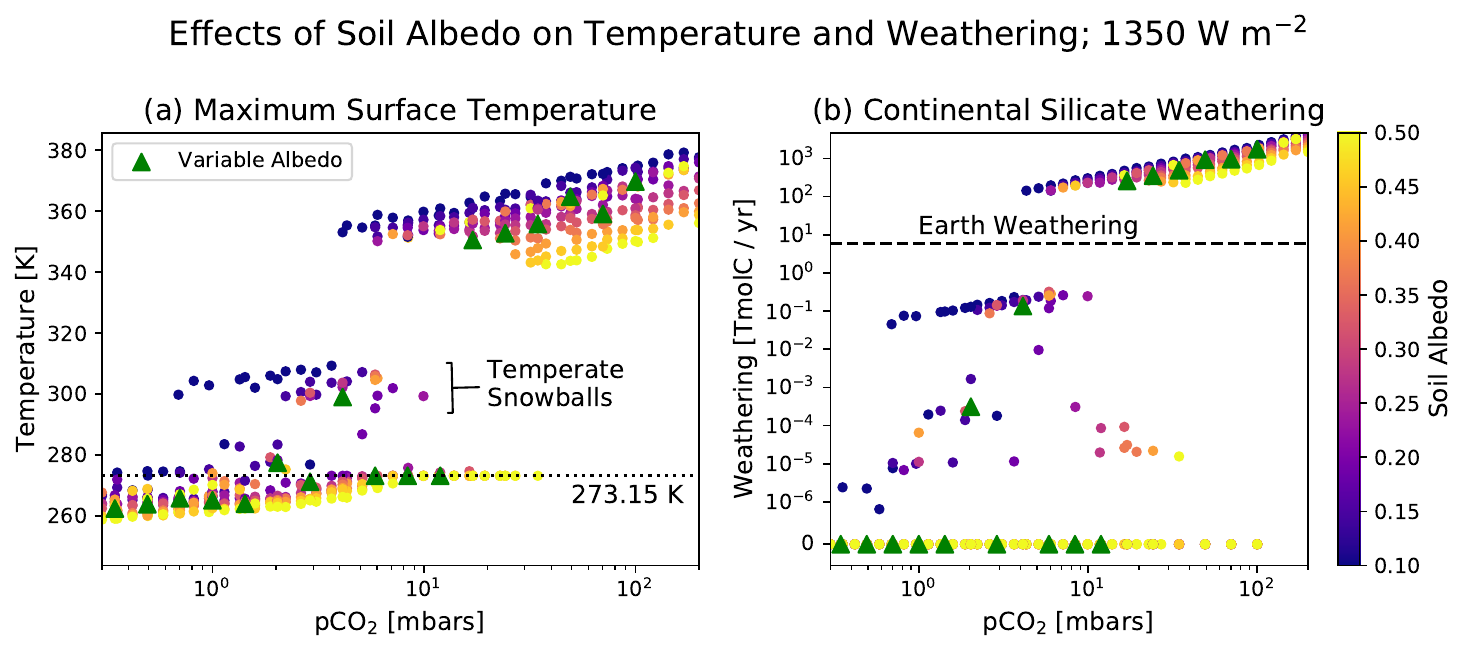}
\end{center}
\caption{Maximum annual surface temperature (\textbf{a}) and net annual CO$_2$ silicate weathering (\textbf{b}) in models with varying pCO$_2$ and soil albedos. The insolation is 1350 W m$^{-2}$. The green triangles indicate models where soil albedo was allowed to vary with moisture content according to \autoref{eq:albedo}. Low albedos produce temperate areas more easily and at lower pCO$_2$, and albedos above 0.4 or so (similar to ice) do not seem to produce temperate snowballs. The transitions from cold snowball to temperate snowball to globally-warm are not as cleanly-separated as in panel \textbf{(b)} of \autoref{fig:parameterspacetemps} in some cases likely because these models were run to temperature equilibrium rather than strict energy balance equilibrium.}
\label{fig:albedos}
\end{figure}

The ability of land surface to reach temperate conditions depends most strongly on shortwave forcing, due to the lower albedo of bare ground relative to snow and ice. We confirm this by adjusting the shortwave albedo of bare soil in the model, ranging from 0.1 to 0.5. PlaSim's glacial ice albedo is 0.8, while sea ice is 0.6, and the snow albedo ranges from 0.4 to 0.8 depending on temperature. Our maximum soil albedo is therefore similar to the surrounding snow. We then perform a parameter sweep of pCO$_2$ at the warm end of the snowball regime, using an insolation of 1350 W m$^{-2}$, running models only until the surface temperature changes slower than a given threshold to save computation time. As shown in \autoref{fig:albedos}, we find that temperate areas form more readily if bare soil has a low albedo, and are not observed in our experiment as the soil albedo approaches that of snow and ice. 

\begin{figure}
\begin{center}
\includegraphics[width=4in]{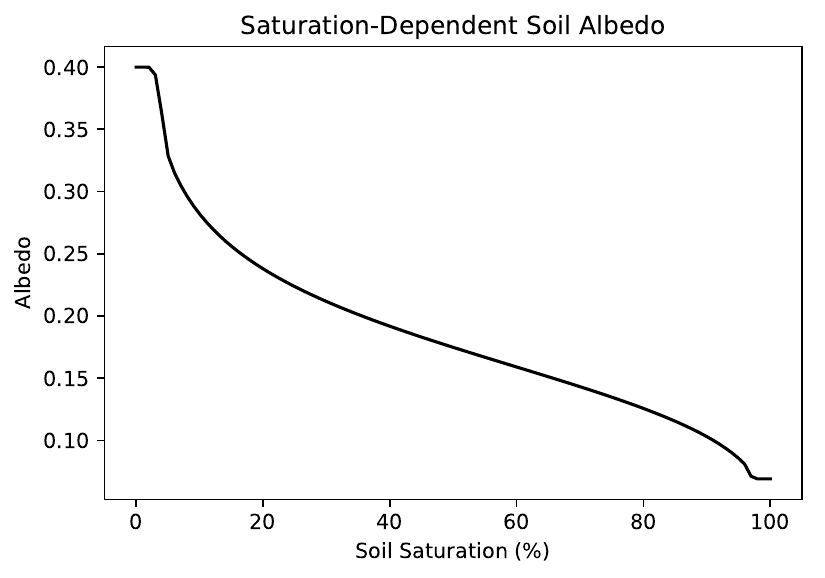}
\end{center}
\caption{The moisture-dependent albedo model we use. This is an ad-hoc qualitative model intended to roughly match \citet{Nolet2014}, such that the albedo converges to dry and saturated values, and darkens quickly at low saturations.}
\label{fig:wetalbedo}
\end{figure}

It is possible that the inclusion of an albedo dependent on soil moisture would inhibit deglaciation---snowball climates have reduced precipitation, and dry ground has a higher albedo. We examine this possibility by constructing an ad-hoc model of soil reflectance as a function of saturation. \citet{Nolet2014} measured the reflectance of beach sand at various saturation levels, and constructed analytical parameterizations to fit the data at various wavelengths. The use of beach sand is probably applicable to the surfaces in our model, since we are assuming that vegetation is not widespread---any surface other than rock is more likely to be primarily sand rather than humus. Our model is based on theirs, and is tuned to qualitatively fit the reflectance behavior described in \citet{Nolet2014}. The moisture-dependence of the soil albedo can be defined as
\begin{linenomath*}
\begin{equation}
A_0=\frac{1}{a}\left(w^\frac{1-y}{y}-1\right)^\frac{1}{y}
\end{equation}
\end{linenomath*}
where $a$ and $y$ are fit parameters, and $w$ is the soil saturation fraction (ranging from 0 to 1). We use $a$=5.2 and $y$=4. To ensure that this function converges to PlaSim's dry soil and ocean albedos at the respective limits of zero water and inundation, we use exponentials with large exponents to smooth $A_0$ into each limit, such that our soil-dependent albedo is parameterized by 
\begin{linenomath*}
\begin{equation}
A = A_de^{-(25w)^6}+A_0\left(1-e^{-(25w)^6}-e^{-[30(1-w)]^9}\right)+A_se^{-[30(1-w)]^9}
\end{equation}
\end{linenomath*}
where $A_d$ is the albedo of dry soil, and $A_s$ is the albedo of saturated soil. We use $A_s$=0.069 and $A_d$=0.4, representing the albedos of the ocean in PlaSim \citep{Fraedrich2005} and the Sahara \citep{Tetzlaff1983}, respectively. This function (shown in \autoref{fig:wetalbedo}), which has the same qualitative shape as the reflectance models in \citet{Nolet2014}, converges to the dry and saturated soil albedos, and darkens significantly at relatively low saturations. Our model spans saturations ranging from 0 to 100\% while the \citet{Nolet2014} models span 0 to 30\% because \citet{Nolet2014} defines saturation fraction as the volume ratio of water to sand, while we mean saturation to be the fraction of the soil bucket's water capacity that is currently filled. \citet{Nolet2014} note that their soil model ceases to be physically-applicable above a volume ratio of 30\%, so we treat that limit as the water capacity in their model. We ignore the reduced liquid water capacity of permafrost, as we assume its effect on this experiment is small. It should be noted that our model and its tunings are primarily intended as an illustrative exploration of the effect of a moisture-dependent albedo on the prevalence of temperate areas in snowball climates, and more realistic models of dynamic soil albedo can be constructed. However, it does provide a semi-realistic way to test whether a moisture-dependent soil albedo affects our results. The maximum annually-averaged surface temperature and annual CO$_2$ weathering observed in models at 1350 W m$^{-2}$ with moisture-dependent albedos are shown as the green triangles in \autoref{fig:albedos}. We find that the case with a variable albedo is qualitatively similar to the cases with a uniform bare soil albedo of 0.2 in warm climates and 0.4 in cold climates.

\subsection{Effect of Glaciology}

Our results depend on the presence and lower albedo of unfrozen, bare ground, which requires that snowpack be melted away entirely. While PlaSim's radiation model includes the reflectivity of ice and snow, and the surface and hydrological models include the various heat capacities and phase transition energies of water, PlaSim lacks a dynamic ice sheet model and neglects glacial surface elevation \citep{Fraedrich2005}. We explore whether this could affect our results by extending PlaSim with a simplistic mass-balance and elevation toy model (as in, we use changes in a cell's ice mass to change the elevation of the ice surface) to incorporate the effects of snow/ice accumulation and elevation on geological timescales. 

We account for dynamic ice sheet growth and collapse by extrapolating the 3-year annual average change in snow depth at each cell, advancing the ice model forward in time such that the maximum change in depth is no more than 300 meters of liquid water equivalent. We use this quantity rather than actual thickness because PlaSim only tracks the ice/snow mass in terms of its liquid water equivalent depth. We convert the snow depth in liquid water equivalent to glacier height by assuming an average glacial density of 850 kg m$^{-3}$ and neglecting spreading and deformation. Noting that spreading and deformation would realistically limit ice sheet height, we arbitrarily limit the height to 3.5 km. We use this new additional elevation to recompute the surface geopotential height. The geopotential height includes both contributions from topography and ice sheets, so that the total elevation over mountain ranges is greater than over basins. We then allow the GCM to relax to the new conditions. Within PlaSim, a cell's surface type is permitted to change from glacial to non-glacial and vice versa throughout the course of the simulation. If a cell has managed to continually retain at least 2 meters of snow in liquid water equivalent for a full year or has reached a snow depth equivalent to 30 meters of liquid water, we treat that cell's surface type as glacial for the following simulation year. This set of tunings is ad-hoc, but we feel they are sufficiently reasonable that our model provides an illustrative test case for continental glacier growth and decay. It should be stressed that this is not a realistic ice sheet or glacier model, but can be useful for assessing the probability of ice sheet collapse on geological timescales in snowball episodes.

\begin{figure}
\centering
\includegraphics[width=4in]{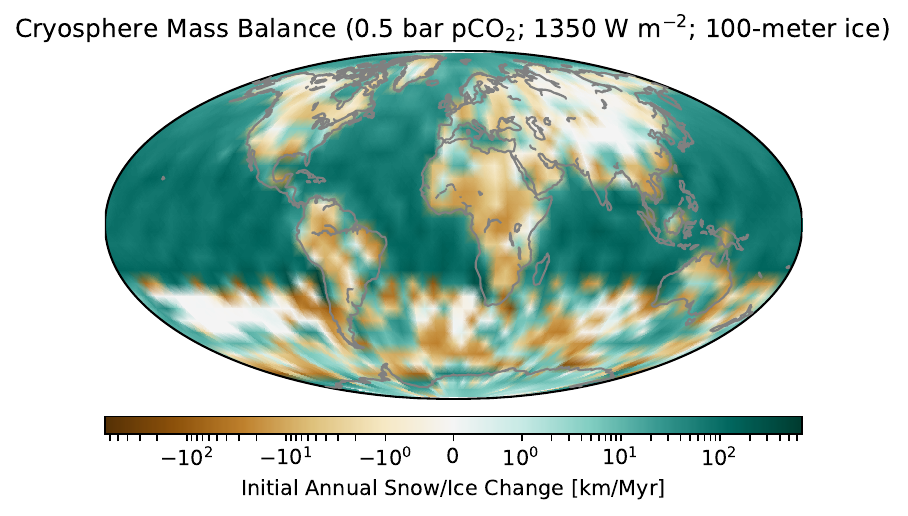}
\caption{Annual average change in thickness of ice sheets with an initial thickness of 100 meters of liquid water equivalent, prior to computing the evolved thickness. This corresponds in our model to an actual thickness of approximately 118 meters, assuming 850 kg m$^{-3}$. This change is computed from the precipitation mass balance. Sea ice has positive mass balance, but sea ice thickness is limited to 10 meters in PlaSim.}
\label{fig:massbalance}
\end{figure}

\begin{figure}
\centering
\includegraphics[width=4in]{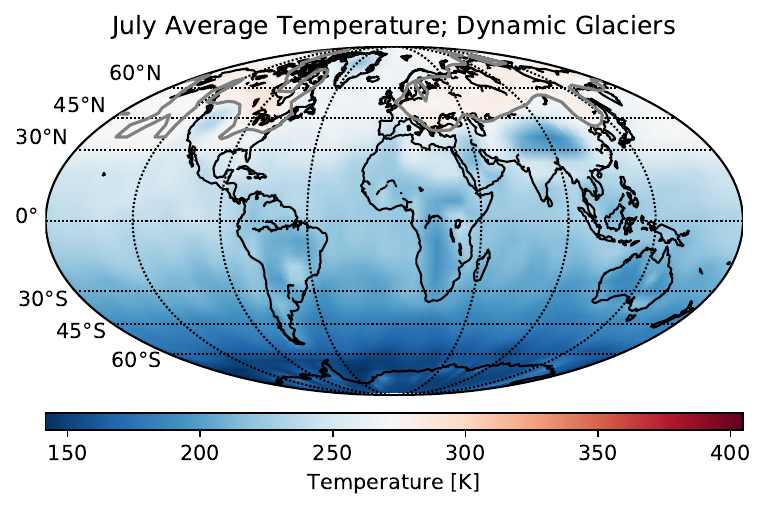}
\caption{Surface temperatures at the warmer end of the snowball regime with evolved initially-thick ice sheets. The temperate areas are smaller and depend more on surface topography, such that elevated areas retain their glaciers. This model also requires more CO$_2$ than the thin-ice model to attain the same temperatures, such that in this case approximately 0.5 bars of CO$_2$ are required.}
\label{fig:glaciers}
\end{figure}

\begin{figure}
\begin{center}
\includegraphics[width=4in]{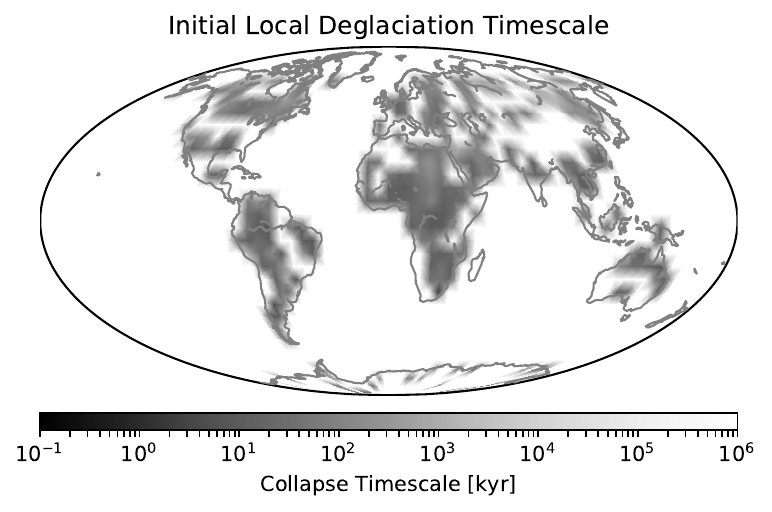}
\end{center}
\caption{The timescale on which an ice sheet will locally collapse at a CO$_2$ partial pressure of 0.5 bars, an insolation of 1350 W m$^{-2}$, and a uniform initial ice thickness of 100 meters. This timescale comes from dividing the thickness by the rate of mass loss. Local elevation includes both ice thickness and the underlying topography. The inclusion of ice sheet elevation reduces the area that deglaciates, but substantial areas nevertheless deglaciate on timescales of 10-100 kyr. For comparison, the Snowball Earth episodes lasted from tens of Myr (Marinoan and Sturtian glaciations) \citep{hoffman98,Rooney2015,Hoffman2017} to hundreds of Myr (the Huronian glaciation) \citep{Kopp2005}. This model remains in a snowball state even once the glaciers have collapsed.}
\label{fig:collapsetime}
\end{figure}

We find that when all land surface in a snowball climate is initially covered by ice sheets that are roughly 1.5 km in thickness, an excess of melting and sublimation relative to snowfall leads the ice sheets to partially collapse on timescales of thousands to tens of thousands of years, resulting in deglaciated temperate land regions. The pre-collapse mass balance and post-collapse surface temperatures are shown in \autoref{fig:massbalance} and \autoref{fig:glaciers}. These regions are in many cases smaller than in the case with thin-ice glaciology, but still span more than one grid cell. Regions of higher average elevation such as mountain ranges do not deglaciate, and neither does most of the land in the Southern Hemisphere. We attribute this to the lower southern hemispheric land fraction. Furthermore, we find that ice sheets initially thicker than 1.5 km do not collapse, resulting in globally cold conditions. We similarly find that scaling the topographic height without ice sheets can increase or reduce temperate areas by raising low-lying areas above the snowline. Therefore, our results are sensitive to the glacial history and dynamics of the planet, as well as its topography. Furthermore, because the rate of glacial collapse is slower for thicker ice sheets, there could be some planets on which warming from CO$_2$ outgassing outpaces glacial collapse, causing sea ice to melt before bare land can be exposed, thereby ending the snowball episode without exposing any ground for weathering. This would result in limit cycles \citep[cf. e.g.][]{Menou2015,HaqqMisra2016,Paradise2017}. The local collapse timescale for a planet with thick-ice glaciology is shown in \autoref{fig:collapsetime}. Our findings that continental weathering could trap planets in snowballs therefore only applies to snowball planets that either do not completely glaciate the land surface or only build ice sheets of low to moderate height.

We do not include changes in sea level in our model, so the planet's water inventory is not conserved. However, 1.5 km of ice on all land points is roughly 10\% of the total mass of Earth's oceans, suggesting that extreme land glaciation would either require a small land fraction or very deep oceans. Planets with larger water inventories and shallower topological relief might therefore be more susceptible to building massive ice sheets that threaten habitability.

Our results are consistent with more robust investigations of the Marinoan snowball Earth in \citet{Benn2015}, in which a 3D ice sheet model (GRISLI) was coupled to a sophisticated GCM (LMDz). In an extension of that work, \citet{Hoffman2017} found the de-glaciated areas in \citet{Benn2015} were able to occasionally reach above-freezing temperatures. In contrast, using different models, \citet{Donnadieu2003} and \citet{Rodehacke2013} were unable to produce deglaciated land in other Snowball Earth experiments, suggesting that the details of the ice sheet model matter, and are one of the key caveats of our results. Our goal was to use a simple approach to identify the impact that ice sheet elevation could have on our results, and find that very large ice sheets are a problem, but thinner ice sheets may not be. There are certainly details of glacial dynamics not captured by our model, and the exploration of their impacts on glacial collapse during snowball events are left for future studies.


\subsection{Soil Hydrology and Erosion}

We also explore the possibility that geological processes such as groundwater, runoff, and erosion might affect our results. These results depend on the presence of liquid water through a local water cycle. PlaSim uses a bucket hydrology for the soil, where each cell has a `bucket' that can hold a prescribed amount of water, with excess treated as runoff. We vary the water capacity of the soil from zero to infinity and find that while the spatial distribution of rainfall and evaporation changes, the weathering rate varies by less than an order of magnitude, such that our qualitative result stands. This is unsurprising, since on large scales evaporation and reaction rates approximately depend primarily on average temperatures through the Arrhenius and Clausius--Clapeyron equations \citep{Berner1994}. The spatial distribution of evaporation and precipitation should therefore be a minor effect so long as too much water vapor is not transported away from the warm evaporative regions.

Weathering also requires fresh rock, which is supplied by geological processes such as uplifting, volcanic eruptions, and erosion. If erosion rates during snowball climates are low, the availability of unweathered rock could be a limiting factor \citep{West2012,Foley2015}. We test our results' sensitivity to this by implementing a supply-limited weathering scheme similar to that proposed in \citet{West2012} and \citet{Foley2015}, where the weathering rate asymptotically approaches a supply limit set by the erosion rate. In this parameterization, supply-limited weathering is given as
\begin{linenomath*}
\begin{equation}\label{eq:supply}
W_\text{sl} = W_\text{max}\left(1-e^{-W/W_\text{max}}\right)
\end{equation}
\end{linenomath*}
where $W_\text{max}$ is the maximum weathering possible given a supply limit, in this case assumed to be set by the erosion rate. $W$ is the weathering that would happen without a supply limit. Weathering rates therefore follow the unlimited formulation when $W/W_\text{max}\ll{1}$, and assume the supply limit when $W/W_\text{max}\gg{1}$. We assume a globally homogenous erosion rate $E$, given in units of length over time. Following \citet{Foley2015}, we define $W_\text{max}$ as
\begin{linenomath*}
\begin{equation}
W_\text{max}=f_l\,E\chi_\text{cc}\,\rho_\text{r}g\frac{\bar{m}_\text{CO2}}{\bar{m}_\text{cc}}
\end{equation}\label{eq:albedo}
\end{linenomath*}
where $f_l$ is the land fraction, $g$ is the surface gravity (9.81 m s$^{-2}$), $\chi_\text{cc}$ is the fraction of Mg, Ca, K, and Na in continental crust (taken to be 0.08), $\rho_\text{r}$ is the density of regolith (taken to be 2500 kg m$^{-3}$), $\bar{m}_\text{CO2}$ is the molar mass of CO$_2$ (44 g mol$^{-1}$), and $\bar{m}_\text{cc}$ is the average molar mass of Mg, Ca, K, and Na (32 g mol$^{-1}$) \citep{West2012,Foley2015}. To determine the impact of limited erosion, we repeat our models of the snowball exit threshold with this modified weathering scheme for erosion rates ranging from 10 cm yr$^{-1}$ to 1 nm yr$^{-1}$. We ignore any potential feedbacks between groundwater runoff and erosion rates, and assume therefore that their limiting effects are not additive: changes in soil hydrology do not affect the magnitude of the supply limit, or vice versa.

\begin{figure}
\begin{center}
\includegraphics[width=4in]{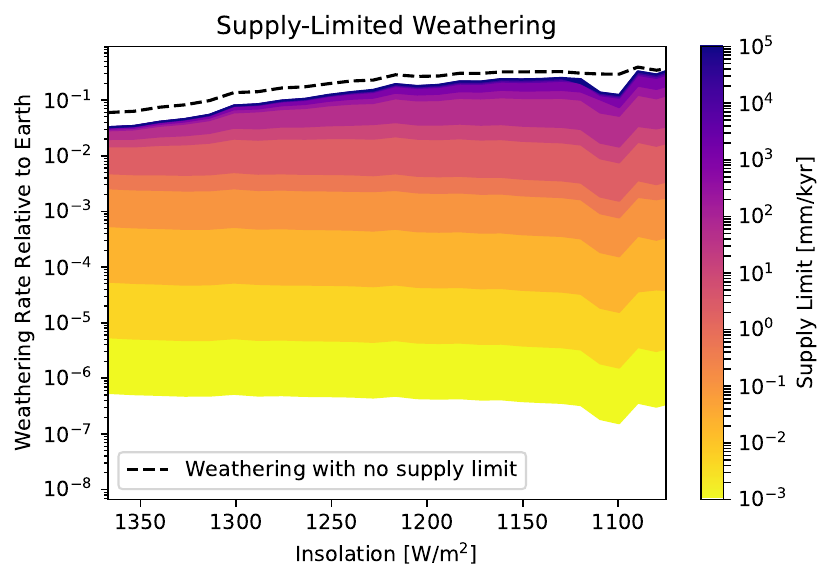}
\end{center}
\caption{The maximum snowball weathering as a function of insolation for planets with varying erosion rates. The kinetic case (no supply limit) is plotted as a dashed line. The difference between this line and the most generous supply limit tested indicates that even a generous limit can reduce overall weathering.}
\label{fig:erosion}
\end{figure}

Our results suggest that an erosion-based supply limit can reduce weathering rates by several orders of magnitude in low-erosion regimes, and a factor of a few in high-erosion regimes, as shown in \autoref{fig:erosion}. The weathering in our models is characterized by very strong weathering in localized areas, so even a generous supply limit can reduce overall weathering. Erosion is primarily controlled by average local slope rather than precipitation \citep{Blanckenburg2005,Willenbring2013}, and glacial activity can dramatically increase erosion \citep{Smith2007}, so the dryness of snowball climates may not reduce erosion below modern rates. Therefore, we do not identify any specific erosion rate as more likely, and view the problem instead as unconstrained in this context.

For perspective, the Antarctic Dry Valleys can experience erosion as low as 25 $\mu$m kyr$^{-1}$ \citep[a limit of 0.2 bar pCO$_2$ Gyr$^{-1}$ if generalized to the global land surface,][]{Schafer1999}, alpine bare bedrock may experience 1 mm kyr$^{-1}$ \citep[8 bar pCO$_2$ Gyr$^{-1}$,][]{Small1997}, and average rates globally may be 0.1--1 m kyr$^{-1}$ \citep[$10^2$--$10^3$ bar pCO$_2$ Gyr$^{-1}$,][]{Willenbring2013}, far in excess of our assumed modern weathering rate. Our results suggest that assuming any of these regimes as representative of snowball erosion can lead to vastly different global weathering rates, which makes an understanding of erosive processes during a snowball event crucial to understanding limits on continental weathering in temperate regions. We do however note that an older and less tectonically-active planet might have less orogeny and thus less erosion, potentially providing a planetary age-based constraint. As a corollary, young snowballs may be more common on active planets if snowball episodes are only long-lived on inactive planets.

\subsection{Other CO$_2$ Sinks}

Our analysis has focused on the continental CO$_2$ weathering permitted by temperate conditions; however this is not likely to be the only CO$_2$ sink active during a snowball climate. Seafloor weathering has a poorly-constrained functional form \citep{Caldeira1995,Abbot2012a}, but may have a much weaker dependence on the surface temperature than surface silicate weathering. If seafloor weathering is sufficiently vigorous and outgassing sufficiently slow, then a snowball planet could reach weathering equilibrium at lower CO$_2$ partial pressures than those required for temperate conditions \citep{LeHir2008}. Similarly, sub-glacial weathering caused by pressure-induced basal melting provides a source of CO$_2$ weathering even when the land surface is buried under thick ice sheets \citep{Boulton2006,Wadham2010}. CO$_2$ condensation could also provide a sink for CO$_2$ and inhibit buildup \citep{Turbet2017,Levi2017}. All of the snowball models considered in our work exhibit minimum temperatures below the CO$_2$ freezing point for that model's CO$_2$ partial pressure, but only in small parts of the polar regions and only during local winter. We also note as in \citet{Paradise2017} that PlaSim exhibits cooling biases at the elevated pCO$_2$ levels of snowball climates, so we are unable to say whether CO$_2$ condensation poses a problem for our models specifically.

\subsection{Other Model Caveats}

Our results include a number of caveats, primarily related to model limitations. For example, PlaSim does not account for changes in sea level as the mass of the ice sheets changes. This would tend to increase continental weathering due to the larger land area. Our results further out in the habitable zone are also less certain than those at higher insolations, as the pCO$_2$ levels associated with those climates are of order 1 bar or higher. As shown in \citet{Paradise2017}, PlaSim performs poorly in the regime of high-CO$_2$ atmospheres, demonstrating significant cooling biases in outgoing longwave radiation and leaving out relevant processes such as CO$_2$ cloud condensation \citep{Forget1997,Kitzmann2017}. We also omit the effects of dust and other aerosols, which have been identified as important components of the snowball radiative budget \citep{Schatten1982,Abbot2010a,Goodman2013}. Dust could tend to reduce the albedo and thus increase warming, allowing planets to exit snowball or reach stable locally-temperate states at lower pCO$_2$, thus reducing the continental weathering permitted during snowball \citep{LeHir2010}, but may also increase cooling of non-glaciated areas by raising the top-of-atmosphere albedo \citep{Claquin2003}.

\section{Conclusion}

We find that snowball climates can feature widespread temperate land areas across a range of parameters, including land distribution, insolation, obliquity, and soil water capacity. These climates sustain low to moderate weathering rates, implying potential for long-term stability. Together with the potential for climate cycles consisting of long snowball events \citep{Menou2015,Paradise2017}, our results suggest that some fraction of observed terrestrial habitable-zone planets may be snowball planets. This may be particularly important for older planets with lower outgassing as well as younger planets in the outer habitable zone, where snowball planets become likely even at Earth-like outgassing rates. 

Together with the results in \citet{Abbot2013} suggesting small areas of open ocean, our results expand the prospects of habitability during snowball events. Along with \citet{Linsenmeier2015}, who found that snowball planets with high axial tilt and eccentricity could have seasonally temperate regions, this bolsters the argument in \citet{Spiegel2008} that some terrestrial planets could have fractional habitability, with habitable areas restricted to certain times of the year or certain regions of the planet. 

Moreover, it may be possible to observationally distinguish these planets from temperate planets. In addition to higher average albedos, snowball planets have high pCO$_2$ but low average atmospheric water content, suggesting elevated CO$_2$/H$_2$O ratios. Snowballs with fractional habitability may have CO$_2$/H$_2$O ratios higher than globally temperate planets but lower than completely-frozen snowballs, due to continued weathering and the presence of limited hydrological cycles. 

These results likely do not however generalize to tidally-locked planets---\citet{Checlair2017} and \citet{Abbot2018} found that tidally-locked and slowly-rotating planets do not undergo the same snowball hysteresis that Earth-like planets exhibit. Our finding that bare soil subject to strong insolation can reach warm temperatures even in mostly-frozen climates would apply to substellar land in such cases, but if such planets do not undergo a sharp transition from snowball conditions to globally-warm conditions, then our partially-temperate snowballs would simply be another state on a continuum for tidally-locked planets.

These results also have implications for the climate cycles explored in e.g., \citet{Menou2015}, \citet{Abbot2016}, and \citet{Paradise2017}---the parameter space in which no weathering equilibrium is expected is greatly reduced by our results, and depends much more strongly on factors such as the planet's glaciology, erosive processes, and topography. We also note that at high insolations, the maximum snowball weathering is greater than the minimum weathering supported by temperate climates described by \citet{Abbot2016} and \citet{Paradise2017}. This suggests that for planets with insolations and outgassing rates in that overlapping region, weathering equilibria are possible for both temperate and snowball states, with the outcome dependent on other factors.

\acknowledgments

AP is supported by a Centre for Planetary Sciences Graduate Fellowship at the University of Toronto, Scarborough, by the Lachlan Gilchrist Fellowship, and by the Department of Astronomy \& Astrophysics at the University of Toronto, St. George. KM is supported by the Natural Sciences and Engineering Research Council of Canada. Computing time was provided by the Canadian Institute for Theoretical Astrophysics at the University of Toronto, St. George. CL is supported by the Department of Physics at the University of Toronto. DV is also supported by the Natural Sciences and Engineering Research Council of Canada. We extend particular thanks to Dorian Abbot for his comments and input with regard to ice sheet dynamics and seafloor weathering, and as well to the other two anonymous reviewers, who posed many important questions. We thank Norm Sleep for his extremely valuable insights on glacial albedo feedbacks, climate-ice sheet interactions, erosion rates, and the geophysical histories of Earth and Mars. We would also like to thank Natasha Kataeva and Taylor Werthauser Jeffries for their assistance with the plain-language summary. This paper is theoretical and does not make use of observational datasets; the modifications to PlaSim necessary for reproduction are available on AP's github \citep{gplasim}. Figures and analysis were done with matplotlib \citep{Hunter:2007} and NumPy \citep{numpy}.

We would furthermore like to acknowledge that our work was performed on land traditionally inhabited by the Wendat, the Anishnaabeg, Haudenosaunee, M\'{e}tis, and the Mississaugas of the Credit First Nation.


\end{document}